\documentclass[aps,prd,floatfix,10pt,nofootinbib,showpacs,notitlepage,superscriptaddress]{revtex4-1} 

\newcommand{\eq}[1]{Eq.~(\ref{#1})}
\newcommand{\fig}[1]{fig.~\ref{#1}}

\newcommand{\bsub}{\begin{subequations}}
\newcommand{\esub}{\end{subequations}}
\newcommand{\be}{\begin{eqnarray}}
\newcommand{\ee}{\end{eqnarray}}
\newcommand{\om}{\ensuremath{\omega}}

\newcommand{\xt}{\ensuremath{x_\om^{\rm tp}}} 
\newcommand{\kt}{\ensuremath{k_\om^{\rm tp}}} 
 
\newcommand{\qt}{\ensuremath{K_\om^{\rm tp}}}

\newcommand{\ommin}{\ensuremath{\omega_{\rm min}}}

\newcommand{\pd}{\ensuremath{\partial}}

\newcommand{\lp}{\ensuremath{\left(}}
\newcommand{\rp}{\ensuremath{\right)}}
\newcommand{\bi} {\begin{itemize}}
\newcommand{\ei} {\end{itemize}}
\newcommand{\ben}{\begin{enumerate}}
\newcommand{\een}{\end{enumerate}}
\newcommand{\bmat}{\begin{pmatrix}}
\newcommand{\emat}{\end{pmatrix}}

\usepackage[colorlinks=true,linkcolor=blue,citecolor=blue,urlcolor=blue]{hyperref}
\usepackage{graphicx}
\usepackage{subfig}
\usepackage{amsmath}
\usepackage{slashed}
\usepackage[utf8]{inputenc}   
\usepackage[english]{babel}
\usepackage{color}
\usepackage{amssymb}
\usepackage{mathrsfs}
\usepackage{multido}

\makeatletter
\renewcommand*\env@matrix[1][c]{\hskip -\arraycolsep
  \let\@ifnextchar\new@ifnextchar
  \array{*\c@MaxMatrixCols #1}}
\makeatother

\begin{abstract}
We study and measure the transmission coefficient of counterpropagating shallow-water waves produced by a wave generator and scattered by an obstacle. To precisely compare theoretical predictions and experimental data, we consider $\sim 25$ frequencies for five subcritical background flows, where the maximum value of the Froude number ranges from $0.5$ to $0.75$. For each flow, the transmission coefficient displays a sharp transition separating total transmission from wave blocking. Both the width and the central frequency of the transition are in good agreement with their theoretical values. The shape of the obstacle is identical to that used by the Vancouver team in the recent experiment aiming at detecting the analog of stimulated  Hawking radiation. Our results are compatible with the observations that have been reported. They complete them by establishing that the contribution of the transmission coefficient cannot be neglected for the lower half of the probed frequency range.
\end{abstract}

\date{\today}
\begin{document}

\title{Wave blocking and partial transmission in subcritical flows over an obstacle }


\author{Léo-Paul Euvé}
\email{leo.paul.euve@etu.univ-poitiers.fr}
\affiliation{Institut Pprime, UPR 3346, CNRS-Université de Poitiers-ISAE ENSMA
11 Boulevard Marie et Pierre Curie - Téléport 2, BP 30179, 86962 Futuroscope Cedex, France.}
\author{Florent Michel}
\email{florent.michel@th.u-psud.fr}
\affiliation{Laboratoire de Physique Th\'eorique, CNRS UMR 8627, B\^atiment 210, Universit\'e Paris-Sud 11, 91405 Orsay Cedex, France}
\author{Renaud Parentani}
\email{renaud.parentani@th.u-psud.fr}
\affiliation{Laboratoire de Physique Th\'eorique, CNRS UMR 8627, B\^atiment 210, Universit\'e Paris-Sud 11, 91405 Orsay Cedex, France}
\author{Germain Rousseaux}
\email{germain.rousseaux@univ-poitiers.fr}
\affiliation{Institut Pprime, UPR 3346, CNRS-Université de Poitiers-ISAE ENSMA
11 Boulevard Marie et Pierre Curie - Téléport 2, BP 30179, 86962 Futuroscope Cedex, France.}

\maketitle

\section{Introduction}

Following the original proposal of Unruh~\cite{Unruh81,Unruh95}, according to which it should be possible to use fluids to test the Hawking prediction~\cite{Hawking75} that black holes spontaneously emit a steady thermal flux, it has been suggested in~\cite{Schutzhold-U_2002} that surface waves propagating on top of a water flow in a flume could be used to experimentally implement this idea~\cite{Germain5}. Since then, several experiments have been conducted to observe the conversion of counterpropagating shallow water waves into shorter wave lengths modes which occurs near a blocking point~\cite{Nice,Unruh2010} of a stationary inhomogeneous flow. This process is related to the time reversed of the Hawking one, as the effective space-time metric close to that point is similar to that of a white hole. However, to have a good analogy with black hole physics, the flow should be transcritical~\cite{Finazzi:2011jd,Scott-review,Finazzi:2012ew,Finazzi:2012bn,MP14}. That is, the flow velocity $v$ should cross the speed of low frequency waves $c$. In the hydrodynamic language, the Froude number $F = v/c$ should become larger than 1.

It turns out that this condition leads to experimental difficulties which, to our knowledge, have not yet been overcome for water waves in an analogue gravity context~\cite{Barcelo:2005fc}. In fact, in the experiments~\cite{Nice,Unruh2010}, the values of the Froude number are significantly lower than 1. Therefore, to interpret the observational data of these experiments, and to understand the exact relationships with black hole physics, it is important to theoretically study the scattering in subcritical flows. These remarks motivated the recent work~\cite{MP14} to which we shall often refer.
 One of its main conclusions is the following: because of the subcritical character of the flow, the wave blocking only occurs above a certain frequency which is rather high in the frequency ranges probed in the experiments~\cite{Nice,Unruh2010}. As a result, it is imperative to take into account the transmission coefficient.

From a hydrodynamical point of view, it is well known that shallow water waves can be blocked by a counter current if the velocity of the latter is sufficiently large~\cite{Dingemans,Fabrikant,Germain5}. Because of dispersion effects of surface waves, the flow velocity needed to block a stationary wave is larger when considering waves of smaller frequency. Turning this around, it means that a given stationary flow reaching a maximal subcritical velocity over an obstacle will block a counterpropagating wave if its frequency is higher than a certain critical frequency, that we shall call $\ommin$ as in~\cite{MP14}. The above reasoning, although correct, is too simplistic as it is based on a geometrical (WKB) approximation. When considering the solutions of the wave equation describing the propagation in a flow over a localized obstacle, one finds that counterpropagating waves are partially transmitted irrespectively of their frequency. The contact with the geometrical approximation is made when noticing that, for increasing values of the frequency, the transmission coefficient monotonically decreases from essentially 1 (i.e., total transmission) to zero (i.e., blocking), with a rather sharp transition in a narrow frequency region close to the critical frequency $\ommin$.

The main purpose of this work is to study, both experimentally and theoretically, the behavior of the transmission coefficient as a function of the wave frequency. We shall work with a given obstacle and current, but with five different values of the asymptotic water depth giving values of $\ommin$ which range from $1.6$ to $5.0$Hz. In spite of the experimental uncertainties and the theoretical approximations, we find a relatively good quantitative agreement between the theoretical predictions and the observations, i.e., with relative differences $\lesssim 10 \%$. 

Importantly, this agreement \textit{a posteriori} validates the various approximations, and the procedures that we have used. On the theoretical side, the approximations are twofold. First, we choose a simplified description of the background flow, avoiding the need to solve partial differential equations. Second, we truncate the wave equation to fourth order in the derivatives, i.e., to lowest nontrivial order in the dispersive scale. Besides the standard quartic wave equation, we introduce a new quartic equation with coefficients tuned so that the exact value of the critical frequency $\ommin$ is taken into account. Both schemes are used and compared to experimental data. On the experimental side, we used acoustic sensors to measure the free surface. The set of $\sim 25$ probed frequencies of the incoming wave allows us to observe, for low frequencies, rather sharp peaks in the transition coefficient which seem to be unrelated to the transmission, but due to resonant properties of the flume, probably associated with (multiple) reflections at its ends.

We believe that the quantitative agreement 
between theory and observations we obtain constitutes an important step both from a purely hydrodynamic point of view, and for the analogue gravity program. Indeed, a good quantitative agreement is needed in order to reliably test in media longstanding predictions concerning the behavior of relativistic fields in curved space-times. 

This paper is organized as follows. In Sec.~\ref{Estd}, we present the experimental setup, and the basic elements for describing the mode mixing occurring in a stationary flow over an obstacle. We also present the aforementioned new quartic wave equation which aims to improve the description of dispersion effects near a turning point. In Sec.~\ref{Tder} we first present the theoretical results obtained by numerically solving the standard and the improved wave equations. We then present the experimental results, that we compare with the theoretical ones. We conclude in Sec.~\ref{Con}. In the Appendix, we present our method to obtain an approximate description of the background flows used in the experiment.

\section{Experimental setup and theoretical description} 
\label{Estd}

\subsection{Experimental setup}
\label{subsection_experim_setup}

\begin{figure} [h]
\begin{center}
\includegraphics[width=\linewidth]{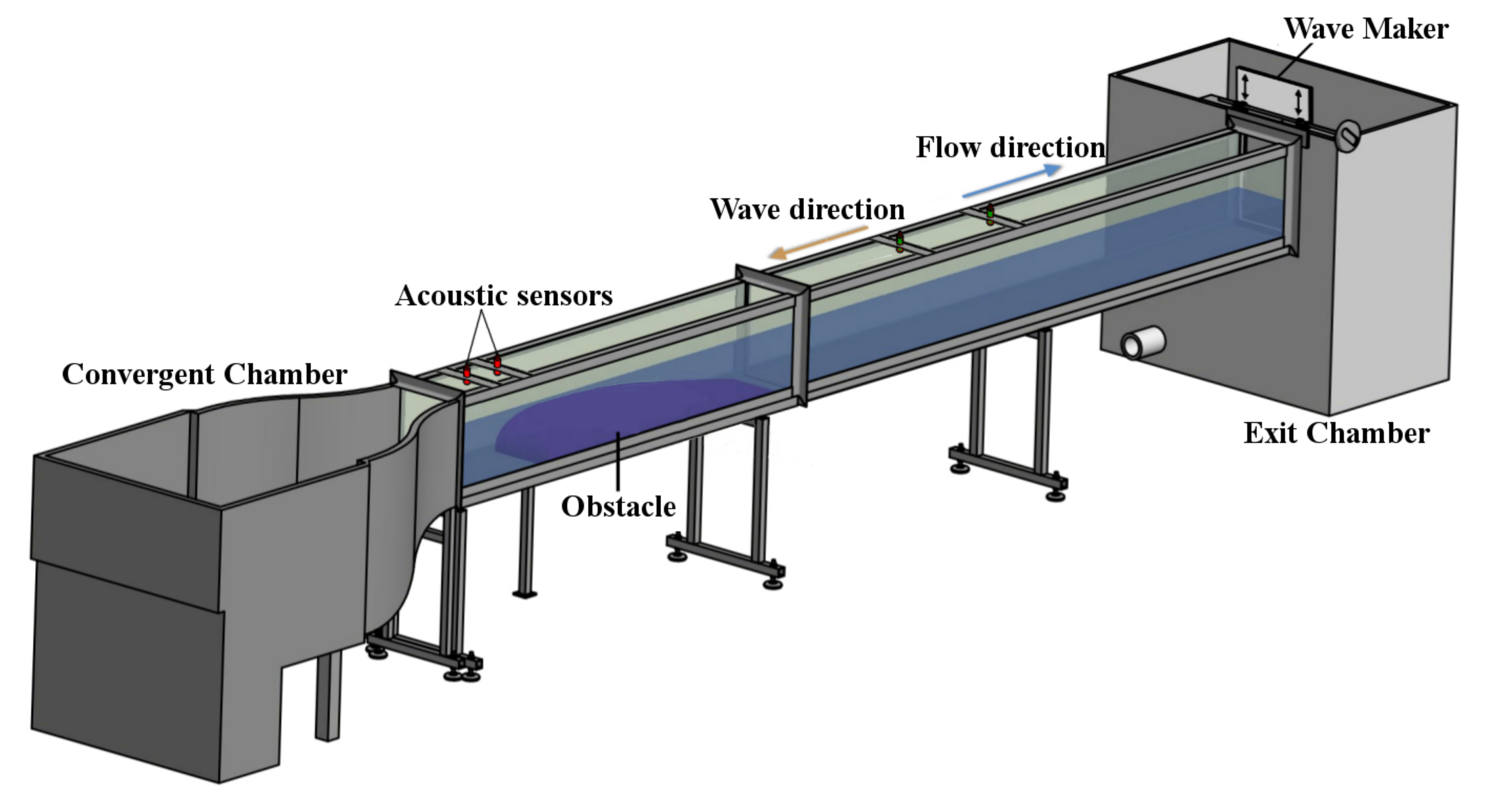}
\end{center}
\caption{A sketch of the experimental setup with the obstacle, the wave making machine, and the four acoustic sensors. The two upstream ones, the red cylinders, measure the amplitude of the transmitted wave on the left of the obstacle. The other two measure the incident wave amplitude.  
}
\label{canal2}
\end{figure}

The experiments were made in a water channel of the Pprime Institute. The flume measures 6.8m of length and 0.39m of width, and the water height can be set up to 0.5m. The current is generated by a PCM Moineau pump with an eccentric rotor bearing (its flow rate per unit width $q=Q/W$ can range up to $0.172m^2/s$). The flow passes through a honeycomb and a convergent chamber to suppress boundary layer effects, macrovortices and turbulence; see~\fig{canal2}. The three-dimensional shape of the convergent chamber is designed to generate a velocity profile at the entrance of the channel which is uniform in the vertical direction. The fluid moves along the water channel of which the sides are made of window panes for visualizations. The fluid is sucked by the pump in the exit chamber at the end of the channel. A gate is placed at its entrance in order to control the exit water depth and flow regimes (typically with a fall at the outlet of the channel). The flow is adjusted at $Q = 10.75 l/s$, corresponding to a flow per unit with $q = 0.0276 m^2/s$.

\begin{figure} [h]
		\begin{center}
		\includegraphics[height=8cm]{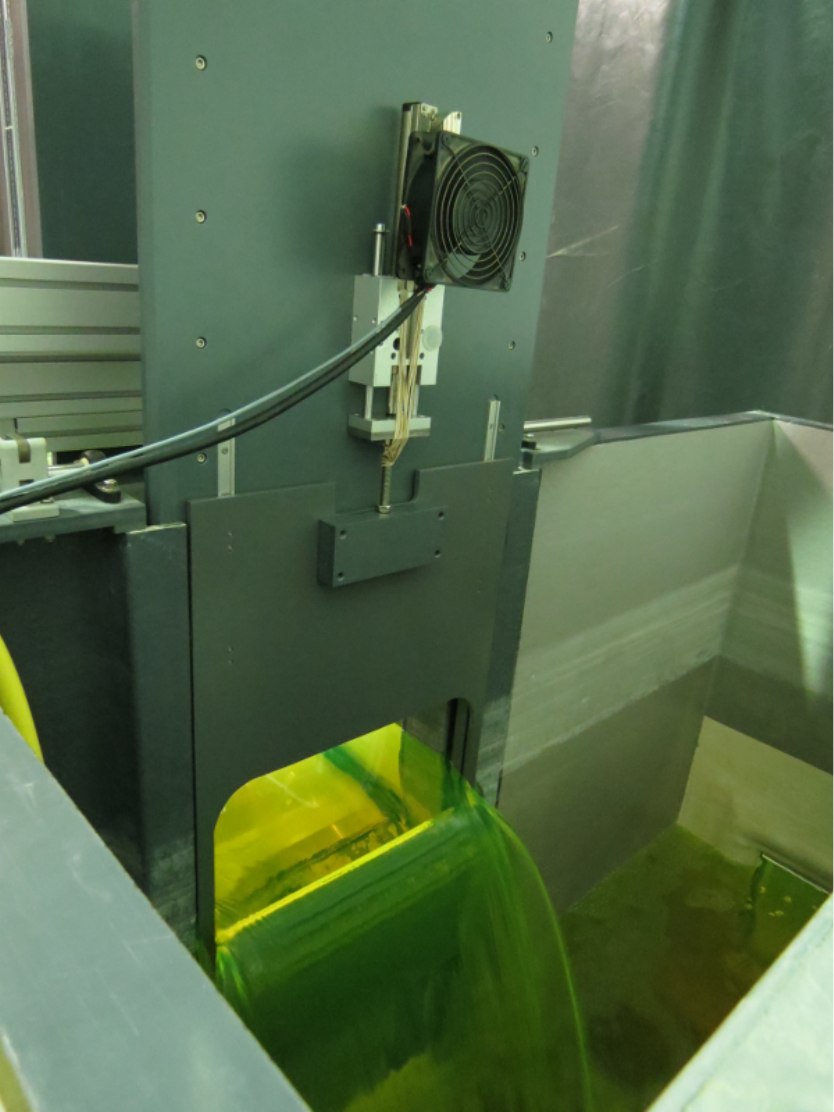}
		\includegraphics[height=8cm]{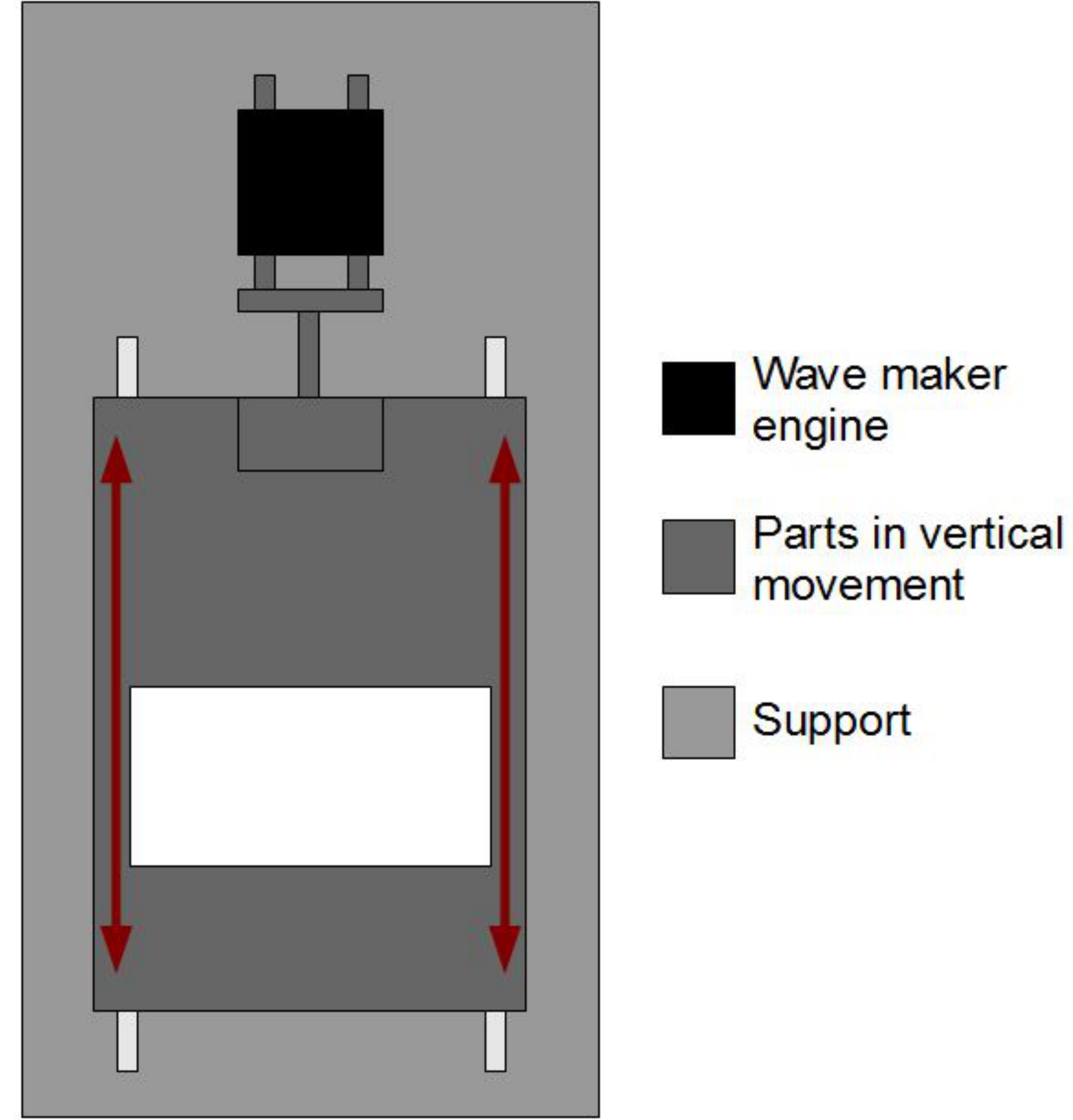}
		\end{center}
		\caption{(Left) The wave-maker at the outlet of the water channel. (Right) A scheme of the mechanical system composing the wave maker.}
		\label{batteur}
\end{figure}

\begin{figure}
\begin{center}
\includegraphics[scale=0.8]{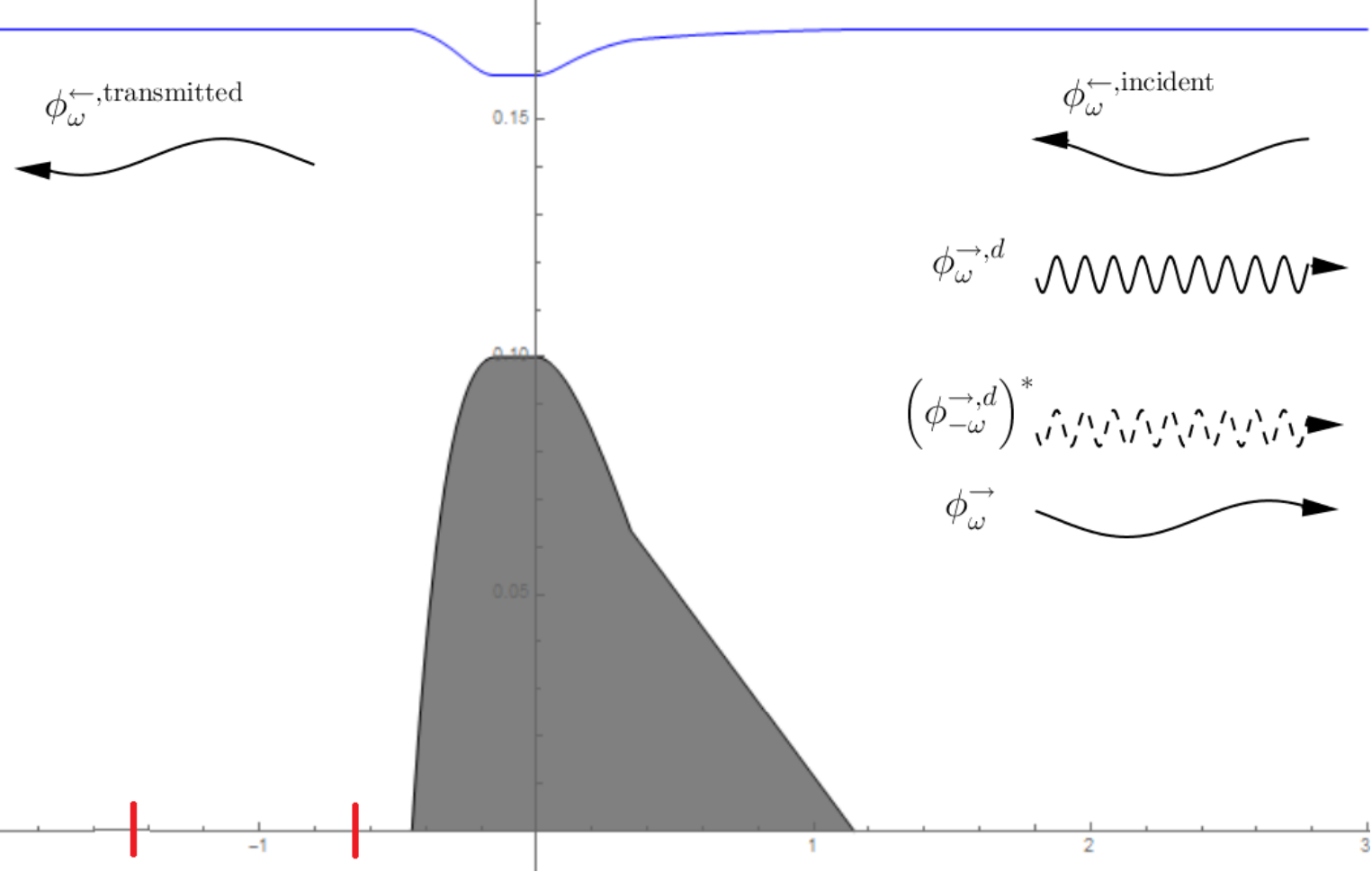}
\end{center}
\caption{Schematic drawing of the obstacle (in gray), the associated flow, and the scattering it induces. The unit of length is the meter. The free surface is represented for the value of the current used in our experiments: $q=0.0276 m^2 \, s^{-1}$, and for an asymptotic water depth equal to 0.169 m, which is the lowest value of the five flows we used, giving the highest value of the Froude number. The flow goes from left to right. The incoming counterpropagating mode generated by the moving guillotine is indicated by the upper arrow on the right side. The transmitted mode is on the left side. The names of the 5 modes are those of \eq{eq:Bsub}. The (horizontal) positions of the two upstream sensors are indicated by the red strokes. }
\label{fig:schema} 
\end{figure}

A wave maker, driven by a linear motor LH23 from Transtechnik (\fig{batteur}), is placed vertically on the gate 
at the end of the channel. It moves (in the vertical direction) a guillotine with a rectangular geometry; see \fig{batteur}. The position of the gate fixes the asymptotic mean water depth in the channel for a given flow rate whereas the motor-driven guillotine superimposes time-dependent perturbations on the mean level, which we set to be periodic. An incoming wave (with a given amplitude for a constant frequency) is thus generated and moves upstream toward the obstacle; see \fig{fig:schema}. Adjusting the position of the exit gate, 5 different asymptotic mean water depths are considered. Their values are $0.169$, $0.1715$, $0.173$, $0.175$, and $0.181m$.

The obstacle we use is the same as in the Vancouver experiments; see Refs \cite{Unruh2010} and \cite{FaltotRousseaux} for more details. Its shape is shown in~\fig{fig:schema}. The leftmost part of the obstacle is placed at 0.8m after the exit of the convergent chamber. 

Two types of measurements of the free surface have been used. First, a laser sheet is created by an Argon LASER (Spectra Physics 2W) connected to a light fiber and a cylindrical lens. It lights the surface perpendicularly and penetrates on a small depth thanks to the absorption of a dye that we diluted in the water channel (see \cite{AGP8} and \cite{FaltotRousseaux}). These measurements shall not be presented here. They have been performed in order to verify that our observations agree with those reported by the Vancouver team. The second type of measurements is made by acoustic sensors (Microsonic mic+25/IU/TC). They measure the water depth as a function of time. They use an echo-radar ultrasound technology with an acquisition frequency of 31 Hz. Their accuracy is 0.1mm. The signal produced is sent to a software developed in the Pprime Institute with Labview. Two pairs of sensors were put on either side of the obstacle to measure the amplitude of the transmitted wave, and that of the incoming one. The first two are placed 0.30m and 0.70m after the convergent chamber, and the last ones 4.42 and 5.42 m after it. 

\subsection{Scattering of water waves on a flow over an obstacle}
\label{sub:gen}


We review the key concepts needed to describe the scattering of surface water waves propagating against an inhomogeneous stationary flow such as that presented above. 
To this end, we first present the simplified wave equation we shall use. Then we study the roots of the dispersion relation 
and the set of stationary modes. Finally we recall the main properties of the mode mixing. These concepts have been recently studied in details in Ref.~\cite{MP14}. Therefore, in what follows, we shall be rather brief. For more details, the interested reader can consult this reference, as well as earlier ones~\cite{Schutzhold-U_2002,Germain4,Coutant_on_Undulations,MacherB/W,Scott-review,ACRPFS}.

\subsubsection{Simplified wave equation}

We consider irrotational, laminar, stationary flows of an inviscid, ideal, incompressible fluid in an elongated flume. The flow profiles are asymptotically uniform on both sides, but possess nontrivial gradients induced by the obstacle put on the bottom of the flume. In the body of the text, we assume that the background flows are known. In Appendix~\ref{app:bf} we explain the method we used to get an approximate analytical description of these flows. The main approximation consists of neglecting the vertical gradients of the flow and the vertical velocity. This is a legitimate approximation in our experiments because the slope of the free surface, which is equal to the ratio of the vertical and horizontal velocities, remains smaller than $10 \%$. 

Using this approximation, the flows are fully characterized by $v(x)$, the velocity in the longitudinal direction $x$, and the water height $h(x)$. Notice that these quantities are related by $v(x) h(x) = q$, where $q$ is the (uniform) value of the current. 
Under this approximation, one easily verifies that the wave equation derived in Appendix A.3 of Ref.~\cite{Coutant_on_Undulations} 
reduces to\footnote{Starting from Eq.~(A25) in Ref.~\cite{Coutant_on_Undulations}, one uses the continuity equation to write $\pd_x v_x = -\pd_y v_y \approx 0$, so that, using Eq.~(A18) in this reference, $v_x G \approx g$. Integrating the definition of the stream function $\Psi$ along the vertical direction gives $\psi_S \approx v_x h$, so that the argument of the hyperbolic tangent becomes, after neglecting $v_y$, $ -i \frac{\psi_S v_x}{v^2} \pd_x \approx -i h \pd_x$.} 
\be \label{eq:waveeq}
\left[ \left(\partial _t+\partial _xv\right)\left(\partial _t+v\partial _x\right) -i g \partial _x \tanh \left(-i h \partial _x\right)\right] \phi = 0,
\ee 
where $g$ is the gravitational acceleration. The field $\phi(t,x)$ is the perturbation of the velocity potential. It is related to the linear variation of the water depth $\delta h$ through 
\be \label{eq:dh}
\delta h(t,x) = -\frac{1}{g}\left(\partial _t+ v \partial _x\right)\phi(t,x).
\ee 
Because the flows we consider are stationary, we can work with (complex) stationary waves $e^{- i\omega t}\phi_\omega(x)$ with fixed laboratory frequency $\om$, which can be taken positive without restriction~\cite{Coutant_on_Undulations}. At fixed $\om$, the spatial part of the waves obeys
\be \label{eq:om}
\left[ 
\left(-i \omega +\partial _xv\right)\left(-i \omega +v\partial_x\right)  
-i g \partial _x \tanh \left(-i h \partial _x\right)
\right]
\phi_\omega 
= 0 .
\ee
Notice that the ordering of $\partial_x$ and the functions $v(x)$ and $h(x)$  has been preserved. The dispersion relation associated with \eq{eq:om} is
\be \label{eq:disprel}
\lp \om - v(x) k \rp^2 = g k \tanh \lp k h(x) \rp ,
\ee
where $k$ is the wave number. \eq{eq:disprel} can be easily obtained from \eq{eq:om} by writing $\phi_\omega = e^{i \int^x k(x') dx'}$ and by 
neglecting the gradients of $h$ and $v$. In the limit $k h \ll 1$, the right-hand side of the dispersion relation becomes $g h\, k^2$, from which one recovers that the speed of long-wavelength perturbations in the fluid frame is $c(x)=\sqrt{g h(x)}$. Combining this with the conservation of the flow, one obtains the local value of the Froude number $F \equiv v/c$: 
\be
F(x) = \frac{q}{\sqrt{g h^3(x)}}. 
\label{Fn}
\ee

\subsubsection{Homogeneous flows}

\begin{figure}
\begin{center}
\includegraphics[width=0.9 \linewidth]{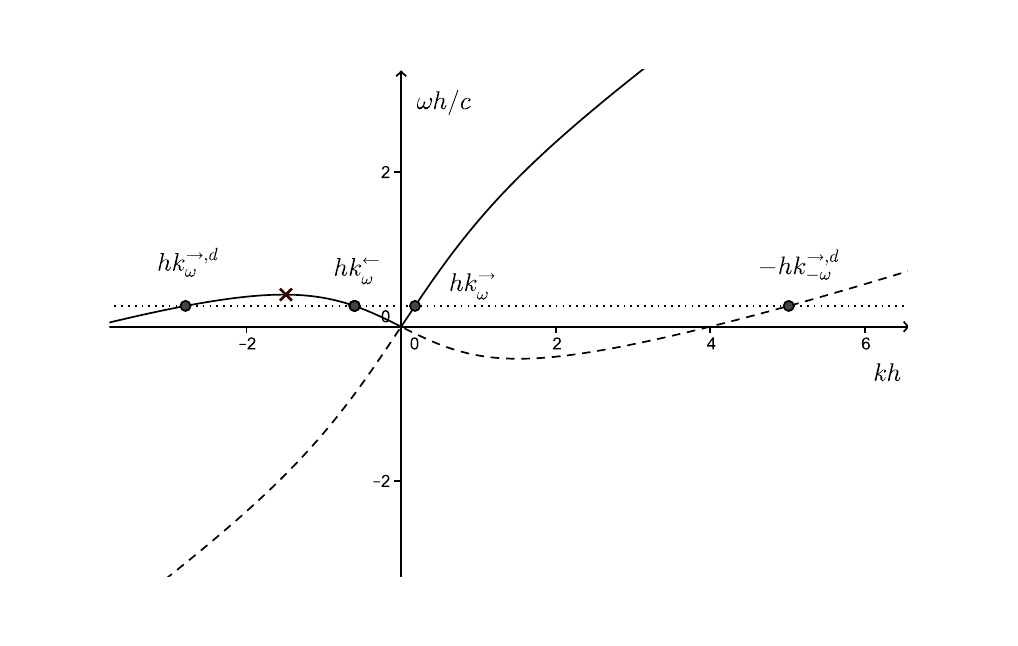}
\end{center}
\caption{Dispersion relation $\om$ vs $k$ [see \eq{eq:disprel}] in nondimensional units, for a Froude number $F=0.5$. The two plain lines represent roots with a positive value of the comoving frequency $\Omega =\omega - v k $, which describe modes with a positive norm of \eq{eq:Kprod}. Instead, the dashed lines describe negative-norm modes with a negative comoving frequency $\Omega$. The dotted line corresponds to $\om = 0.23 c/h < \om_m$. The latter is given by the horizontal tangent on the curve (not represented) on the left upper quadrant, and is indicated by a cross. The dots indicate the four wave vectors for this value of $\om$. Their meaning is explained in the text.} \label{fig:disprel}
\end{figure}

Before studying the mode mixing engendered by the gradients of $h$ and $v$, it is appropriate to study the algebraic properties of the roots of \eq{eq:disprel} in a uniform flow. For definiteness, we assume that the flow velocity $v$ is positive, and smaller than $c$. In subcritical flows, \eq{eq:disprel} has four real solutions if $0< \om < \om_m$, where $\om_m$ is the value of $\om$ where \eq{eq:disprel} has a
real double root; see  \fig{fig:disprel}. When $\om > \om_m$, there are only two real roots. In both cases, \eq{eq:disprel} also has an infinite number of complex roots.

When $0 < \om < \om_m$,  from left to right, the four roots are
\begin{itemize}
\item the dispersive, copropagating root $k_\om^{\rightarrow,d}$;
\item the hydrodynamic, counterpropagating one $k_\om^{\leftarrow}$;
\item the hydrodynamic, copropagating one $k_\om^{\rightarrow}$;
\item the dispersive, copropagating one $-k_{-\om}^{\rightarrow,d}$.
\end{itemize}
As in Ref.~\cite{MP14}, we call a root {\it hydrodynamic} if it vanishes in the limit $\om \to 0$, and {\it dispersive} if it does not. The superscript $d$ is used to distinguish these two types of roots. The arrow gives the sign of the group velocity in the lab frame, and the notation for the last root underlines the fact that the corresponding wave has a {\it negative energy}. To understand this one should describe the plane waves $\phi_\om \propto e^{i k_\om x}$ associated with these four roots. They are respectively called $\phi_\om^{\rightarrow,d}$, $\phi_\om^{\leftarrow}$, $\phi_\om^{\rightarrow}$, and $\lp \phi_{-\om}^{\rightarrow,d} \rp^*$.
The last one, which has a negative energy, is complex conjugated for the following reason. Each of these four (complex) waves possesses a norm which is given by the (conserved) scalar product associated with \eq{eq:waveeq}. Considering two complex solutions $\phi_1, \phi_2$ of \eq{eq:waveeq}, it is given by
\be \label{eq:Kprod}
\lp \phi_1, \phi_2 \rp \equiv i \int \lp \phi_1^* (\pd_t + v \pd_x) \phi_2 - \phi_2 (\pd_t + v \pd_x) \phi_1^* \rp \, dx. 
\ee
One then verifies that the above four modes are orthogonal to each other, which means that they describe independent waves. When working with positive frequency $\omega$, one also finds that the first three modes have both a positive energy and a positive norm. 
Instead, the norm of $\lp \phi_{-\om}^{\rightarrow,d} \rp^*$ is negative. This means that $\phi_{-\om}^{\rightarrow,d}$ has a positive norm and describes negative energy waves, as can be verified by direct evaluation of the wave energy functional [see Ref.~\cite{Coutant_on_Undulations} for details about the relationship between the sign of the norm of the complex modes $e^{- i\om t}\phi_\om$, and the sign of the energy of the corresponding physical waves $Re(e^{- i\om t}\phi_\om)$].
The contact with the experiment is easily made when noticing that the wave generator will send towards the obstacle the counterpropagating mode $\phi_\om^{\leftarrow}$. When scattered near the obstacle, four outgoing waves will be generated, as is schematically represented in \fig{fig:schema}.

When $\om > \om_m$, the two roots $k_\om^{\rightarrow,d}$ and $k_\om^\leftarrow$ become complex and conjugate to each other. The corresponding modes exponentially grow to the left or to the right and do not describe physical waves in homogeneous flows.

\begin{figure}
\begin{center}
\includegraphics[width=0.5 \linewidth]{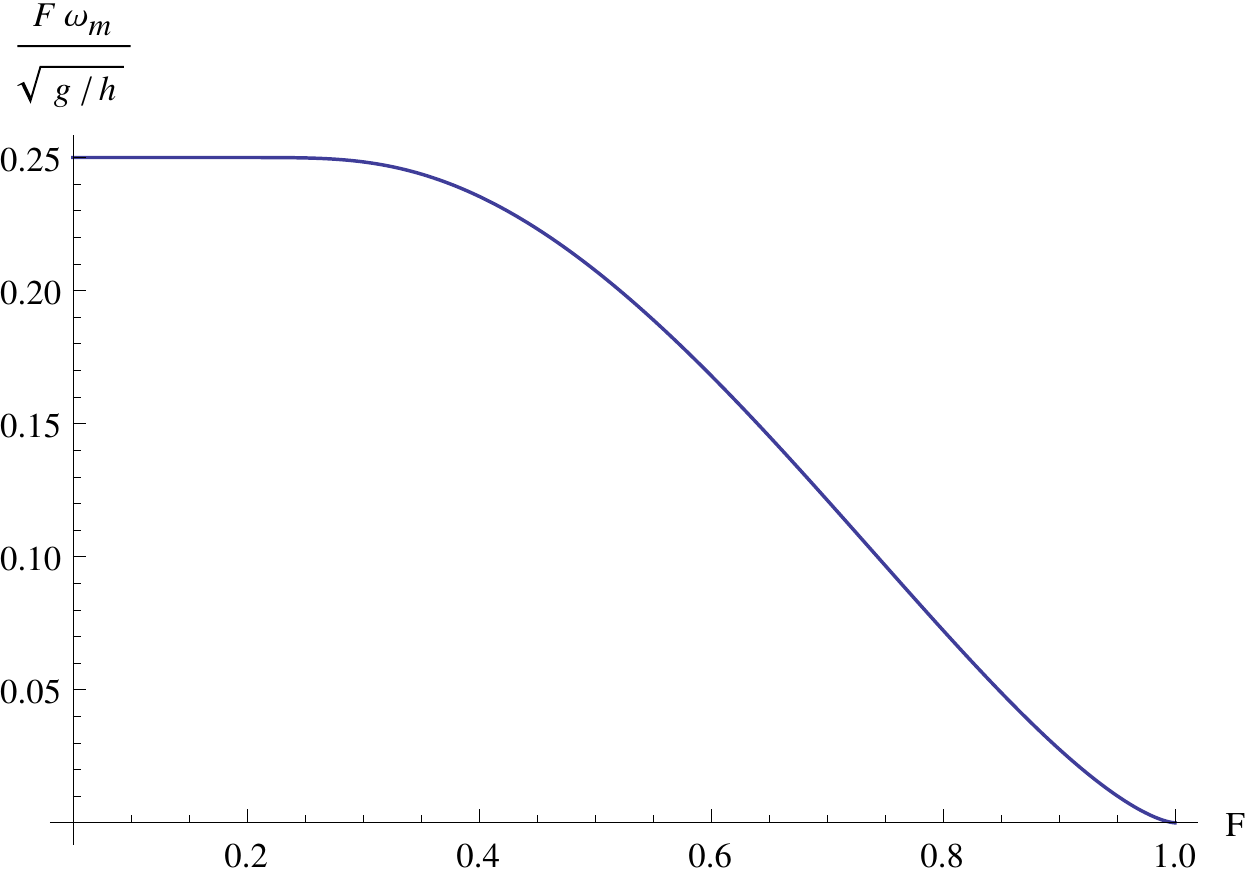}
\end{center}
\caption{Plot of the adimensional critical frequency times the Froude number, $\om_m/ \sqrt{g/h} \times F$ as a function of the Froude number $F$. Notice that $\om_m$ vanishes for $F = 1$, and no longer exists for supercritical flows.} \label{fig:omm}
\end{figure}

To conclude this subsection, it is of value to determine how $\om_m$ depends on the Froude number.  Using the adimensional quantities $\om_m h /c$ and $k_m h$, where $k_m$ is the double root of \eq{eq:disprel} for $\om = \om_m$, we get
\be \label{eq:ommm}
\begin{split}
F &= -\frac{1}{2 \sqrt{k_m h \tanh(k_m h)}} \lp \tanh(k_m h) + k_m h \lp 1-\tanh^2 (k_m h) \rp \rp, \\
\frac{\om_m h}{c} &= \sqrt{k_m h \tanh(k_m h)} + F k_m h. 
\end{split}
\ee
These equations implicitly give $\om_m h /c$ as a function of $F$. The result is shown in \fig{fig:omm}. A straightforward calculation gives its behavior in the two limits $F \to 0$ and $F \to 1$:
\be 
\begin{split}
\frac{\om_m h}{c} & \mathop{\sim}_{F \to 0} \frac{1}{4 |F|}, \\
\frac{\om_m h}{c} & \mathop{\sim}_{F \to 1} \frac{1}{3} \lp 1-F^2 \rp^{3/2}.
\label{eq:omm}
\end{split}
\ee

\subsubsection{Mode mixing in inhomogeneous flows}

When considering inhomogeneous flows in the presence of a localized obstacle, the above analysis should be reconsidered as the four modes will mix with each other. We assume that the flow is asymptotically homogeneous on both sides of the obstacle, with the same values of $h$ and $q$. As a result, the asymptotic properties of the solutions of \eq{eq:om} can be decomposed in plane waves characterized by the four roots of \eq{eq:disprel}. The scattering of the incoming mode $\phi_\om^{\leftarrow, \rm incident}$ from the right is thus {\it completely} described by the four complex coefficients appearing in 
\be 
\label{eq:Bsub}
\phi_\omega^{\leftarrow, \rm incident} \to \alpha_\omega \phi_\om^{\rightarrow,d} + \beta_\omega \lp \phi_{-\om}^{\rightarrow,d} \rp^* + \tilde{A}_\omega \phi_\omega^{\leftarrow, \rm transmitted} + A_\omega \phi_\om^{\rightarrow}.
\ee
A schematic drawing of the scattering is shown in \fig{fig:schema}. Using the conservation of scalar product of \eq{eq:Kprod}, and working with incident and outgoing waves with unit norm, the four coefficients automatically obey the unitarity relation
\be \label{eq:unit}
\left\lvert A_\om \right\rvert^2 + \left\lvert \tilde{A}_\om \right\rvert^2 + \left\lvert \alpha_\om \right\rvert^2 - \left\lvert \beta_\om \right\rvert^2 = 1,
\ee
where the minus sign in front of $\left\lvert \beta_\om \right\rvert^2$ comes from the unit negative norm of $\lp \phi_{-\om}^{\rightarrow,d} \rp^*$. The reader unfamiliar with this important relation can consult Refs.~\cite{MP14,MacherB/W,Scott-review}. In the following we focus on the transmission coefficient $\tilde A_\om$.

In our flows, the mode mixing of \eq{eq:Bsub} heavily depends on two critical frequencies, 
which we call $\om_{\rm max}$ and $\ommin$. These two frequencies are both defined by \eq{eq:omm} and are respectively associated with the minimal and maximal values of the Froude number of \eq{Fn}:
\be
\om_{\rm max} & = \om_m(F = F_{\rm min}), \\
\label{eq:ommax} 
\om_{\rm min} & = \om_m(F = F_{\rm max}).
\label{eq:ommin} 
\ee 
Even though they are defined in a similar manner, they play very different roles, as we now explain. In our flows, $F$ reaches its minimal value in the asymptotic (right) region. Hence $\om_{\rm max}$ is defined by the asymptotic value of \eq{eq:omm} for $x \to \infty$. Since $\om_m$ decreases with $F$, $\om_{\rm max}$ is its largest value. Because no counterpropagating mode exists for $\om > \om_{\rm max}$, the wave generator put on the right of the obstacle can only send waves for $\om < \om_{\rm max}$. Hence, only frequencies {\it lower} than $\om_{\rm max}$ should be considered. 

The second frequency, $\ommin$, is instead defined on top of the obstacle, where $F$ reaches its maximum value $F_{\rm max}$. This frequency plays a crucial role in the mode mixing because for $\om < \om_{\rm min}$, $\phi_\omega^{\leftarrow, \rm incident}$ will be essentially transmitted over the obstacle, since its group velocity (in the fluid frame) is higher than the flow speed. Instead for higher frequencies, $\om > \om_{\rm min}$, the incoming wave $\phi_\omega^{\leftarrow, \rm incident}$ is essentially blocked since it cannot propagate to the left (in a WKB sense) in the region on top of the obstacle where the flow speed is too high. This reflection can be understood from a geometrical optics point of view. When considering the mode characteristics governed by \eq{eq:disprel}, one finds that they posses a turning point at $x = \xt$ where \eq{eq:disprel} admits a double root that we call $\kt$. For more details about this, see Refs~\cite{ACRPFS,MP14}. When working at fixed $\om > \ommin$, the value of $\kt$ is given by $k_m$ of \eq{eq:omm}. For $\om < \ommin$, there is no turning point because \eq{eq:omm} no longer admits a (real) double root. This indicates that the wave is essentially transmitted.
These properties have been theoretically studied in Ref~\cite{MP14} using a combination of analytical and numerical techniques. Our main goal is to refine this analysis and to confront the results with experimental data. 

\subsection{Standard and improved quartic equations} 
\label{sub:num}

Because \eq{eq:om} contains derivatives of arbitrary high orders, standard numerical methods to solve ODEs require some truncation. In this work two truncations shall be used, and their outcomes shall be compared with experimental data. First, as in Ref~\cite{MP14}, we expand \eq{eq:om} to fourth order in $\pd_x$, to arrive at 
\be
\label{eq:stand4ode}
\left[ 
\left(\omega + i\partial _xv\right)\left( \omega + iv\partial_x\right) 
+ g \partial_x 
h \partial_x 
+ \frac{g}{3}\partial_x 
\left(h \partial_x \right)^3 
\right]
\phi_\omega 
= 0 .
\ee
This fourth-order equation can be integrated using standard techniques~\cite{MacherB/W}. However, as the Froude number remains rather far from unity for the flows we shall use, the above expansion around $k=0$ is \textit{a priori} not reliable. Indeed, the relative error between the values of $\ommin$ obtained using the full dispersion relation of \eq{eq:disprel} or the quartic dispersion relation $(\om - vk)^2 = c^2 k ^2 (1 - h^2k^2/3)$ reaches $24 \%$ when $F_{\rm max} = 0.47$ which is the lowest value in the set of five flows we shall study. Notice also that the error on $\om_{\rm max}$ is even larger (close to $50 \%$) since the corresponding values of $F$ are smaller. For intermediate frequencies $\ommin < \om < \om_{\rm max}$, the errors on the location of the turning point $\xt$ and that of the corresponding double root $\kt$ are of the same order since these are also obtained from \eq{eq:omm}.

To avoid these systematic errors, we propose a new method which still consists of expanding the wave equation to fourth order in $\pd_x$, but do so in a $\om$-dependent manner so that, for all frequencies $\ommin < \om < \om_{\rm max}$, the exact value of the turning point $\xt$ and that of the corresponding double root $\kt$ are taken into account. That is, when using the standard WKB approximation to get the effective dispersion relation from the new wave equation, the values of $\xt$ and $\kt$ should coincide with those predicted by the full dispersion relation \eq{eq:disprel}. Since this applies for $\ommin < \om < \om_{\rm max}$, it will also apply for the limiting cases $\ommin$ and $\om_{\rm max}$. Hence, by construction, the new method works with the exact values of $\ommin$ and $\om_{\rm max}$. 

A simple way to proceed consists of taking \eq{eq:stand4ode} and adjusting the coefficients of the last two terms in a $\om$-dependent manner so as to meet the above criterion. To be more specific, we still write the wave equation as
\be 
\left[ 
\left(\omega + i\partial _xv\right)\left( \omega + iv\partial_x\right) 
+ \frac{1}{h} \mathcal{F} (\om, \pd_x) 
\right]
\phi_\om = 0,
\ee
where the factor $1/h(x)$  is introduced for convenience.
Notice that the convective derivative term is the same as that of \eq{eq:waveeq}. In fact, we recover \eq{eq:waveeq} when setting 
$\mathcal{F} = - i g h \pd_x \tanh (-i h \pd_x)$, while \eq{eq:stand4ode} is obtained  for 
$\mathcal{F} = g \lp (h \pd_x)^2 +(1/3) (h \pd_x)^4 \rp$. As there exist many wave equations which meet the criterion on the values of $\xt$ and $\kt$, we also impose that the new $\mathcal{F}$ operator obeys the following properties, which are satisfied both by \eq{eq:waveeq} and \eq{eq:stand4ode}, 
\begin{itemize}
\item $\mathcal{F} \phi_\om = 0$ if $\pd_x \phi_\om = 0$;
\item $\mathcal{F}$ is even in $\pd_x$;
\item $\mathcal{F}$ depends on $h$ and $\pd_x$ only through $h \pd_x$;
\item $\frac{1}{h} \mathcal{F}$ is Hermitian for the scalar product \eq{eq:Kprod} to be conserved.
\end{itemize}
Assuming that $\mathcal{F}$ is fourth order in $\pd_x$ then gives an equation of the form
\be \label{eq:neweq}
\left[ 
\left(\omega + i\partial _xv\right)\left( \omega + iv\partial_x\right) 
+ g_2  \pd_x  h \pd_x 
+ \frac{g_4}{3} 
\partial_x 
\left(h \partial_x \right)^3 
\right]
\phi_\om = 0,
\ee
where $g_2$ and $g_4$ are two real parameters. Their values are fixed by imposing that the associated dispersion relation 
\be 
(\om - v k)^2 = g_2 h k^2 - \frac{g_4}{3} h^3 k^4,
\label{eqdisprelimp}
\ee
is tangent to the exact one \eq{eq:disprel} at $k = \kt$. (This guarantees that the values of $\xt$ and $\kt$ obtained from \eq{eqdisprelimp} coincide with those derived from \eq{eq:disprel}.) A straightforward calculation gives
\be 
\begin{split}
g_2(\om) = \frac{g}{2} \lp 3 \frac{\tanh(\qt)
}{\qt} - 1 + \tanh^2(\qt) \rp, \\
g_4(\om) = \frac{3 g}{2} \lp \frac{\tanh (\qt)}{(\qt)^3} - \frac{1-\tanh^2(\qt)}{(\qt)^2} \rp,
\end{split}
\ee
where the quantity $\qt = \kt \times h(\xt)$ is the wave number $\kt$ adimensionalized by the water height at the turning point. Notice that $g_2$ and $g_4$ both tend to $g$ in the limit $\qt \to 0$. 

So far we considered cases with a turning point. In the absence of a turning point, for $\om < \om_{\rm min}$, we use the same procedure with $\qt$ replaced by the counterpropagating root $K_\om^{\rm top} = k_\om^{\leftarrow}(x_{\rm top}) \times h(x_{\rm top})$ adimensionalized by $h(x_{\rm top})$, the height where the water depth is minimal. Therefore $g_2$ and $g_4$ are continuous across $\om = \ommin$. 

The main assumption underlying \eq{eq:neweq} is that the scattering mainly occurs near the turning point, when there is one, or near the top of the obstacle, when it is absent. It turns out that this condition is not well satisfied for the obstacle used in the experiments since the beginning of its upstream slope is very steep, as can be seen in \fig{fig:schema}. This may explain why the two methods give very similar results, as we shall see in the next section. Yet, we believe that for smoother, and more symmetrical obstacles, the improved description based on \eq{eq:neweq} should give a more accurate description than that based on \eq{eq:stand4ode}. 
We hope to be able to validate this conjecture in the near future. 

\section{Theoretical predictions and experimental results} 
\label{Tder}

\subsection{Numerical results}

\begin{figure}
\begin{center}
\includegraphics[scale=1.0]{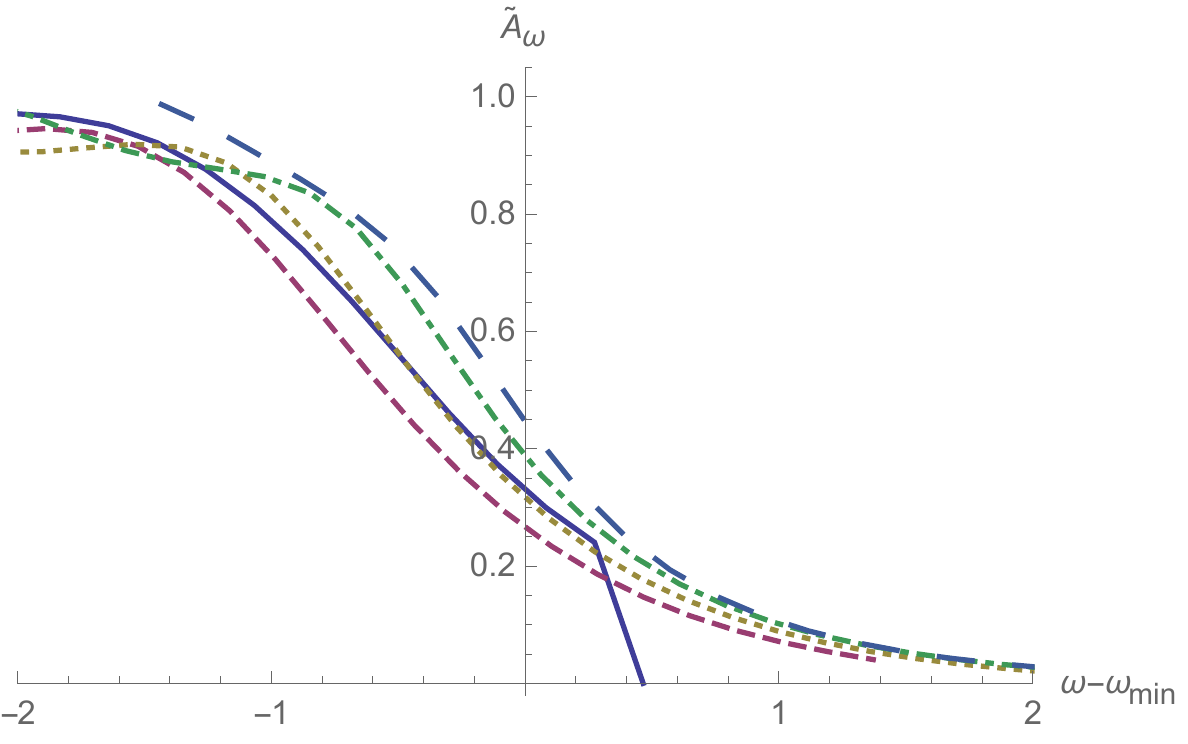}
\includegraphics[scale=1.0]{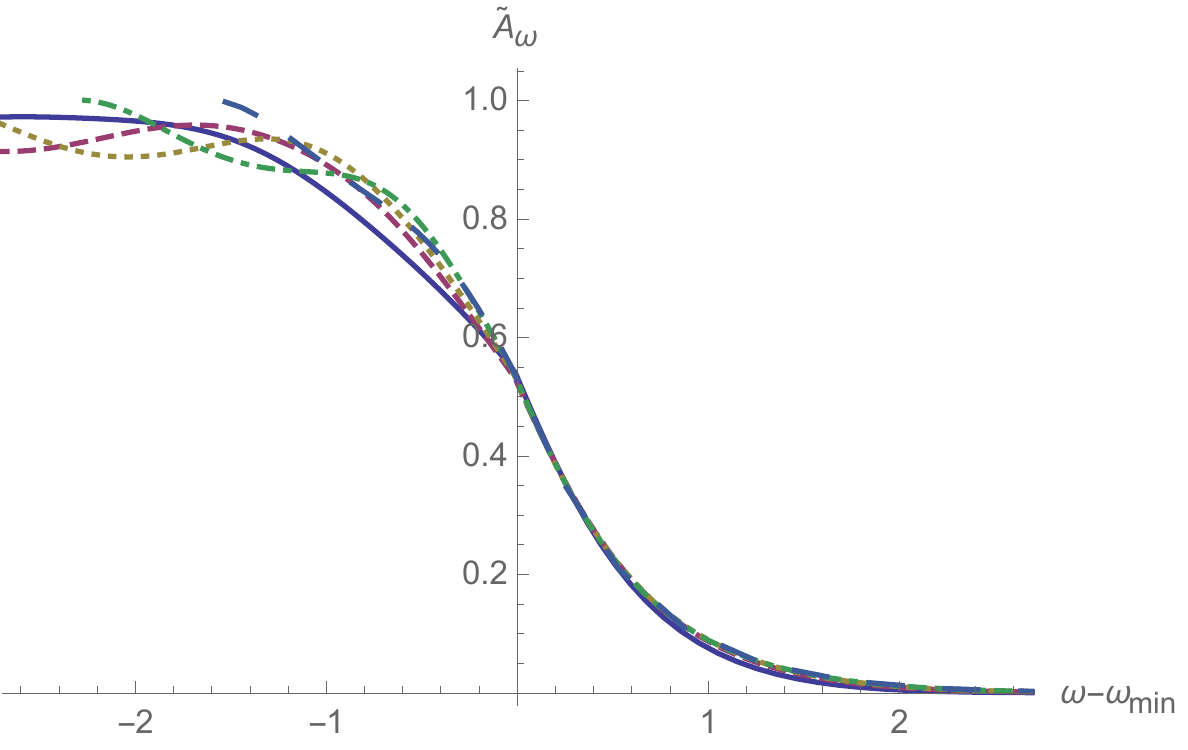}
\caption{Plot of the numerical value of $\tilde{A}_\om$ vs $\om - \ommin$ (expressed in hertz) for the five flows obtained with a fixed value of the current $q = 0.0276 m^2 \cdot s^{-1}$, and five different asymptotic water depths 0.181, 0.175, 0.173, 0.1715, and 0.169 m, for more details see the main text. Top: Using the standard quartic wave equation. Bottom: Using the refined wave equation of \eq{eq:neweq}. The curves in (plain, blue), (purple, dashed), (yellow, dotted), (green, dot-dashed) and (lightblue, long-dashed) correspond to increasing values of the maximal value of the Froude number. The main lesson is that all curves show a sharp decrease from essentially 1 to 0 that occurs in a narrow interval of $\pm 1$Hz, centered at a frequency close to $\ommin$ of \eq{eq:ommin}. Notice that the plain blue curve in the top panel falls off abruptly to $0$ at the value of $\om_{\rm max}$ computed using the naive fourth-order expansion of the dispersion relation.}
\label{fig:numpred}
\end{center}
\end{figure}

We numerically computed the transmission coefficient $\tilde{A}_\om$ by solving the two wave equations of Sec.~\ref{sub:num} for the obstacle used in the Vancouver experiment~\cite{Unruh2010}; see \fig{compVancouver}. We considered five different flows with a fixed current $q=0.0276 m^2 \, s^{-1}$, and water depths at the top of the obstacle given by $h_{\rm min} = 0.0695 $, $0.0615 $, $0.058 $, $0.054 $, and $0.051$ m, corresponding to asymptotic water depths of  0.181, 0.175,
0.173, 0.1715, and 0.169 m, respectively.\footnote{We here use $h_{\rm min}$ rather than the asymptotic water height fixed by the position of the guillotine for the following reasons. First, it is $h_{\rm min}$ which fixes the value of the critical frequency $\omega_{\rm min}$, see Eqs~(\ref{eq:ommm}) and~(\ref{eq:ommin}). Using the asymptotic water depth would require computing $h_{\rm min}$ by a procedure which, in the absence of complete knowledge of the flow, would have introduced an error on $\omega_{\rm min}$. Second, as can be easily seen from Eq.~\ref{eq:nB}, the derivative of $h_{\rm min}$ with respect to the asymptotic water depth is larger than unity. Hence, matching the observed value of $h_{\rm min}$ to the theoretical one automatically gives a more accurate description of the flow. We checked that this choice indeed improves the agreement between theoretical predictions and observations.} The corresponding values of $F_{\rm max}$ and $\ommin$ (in hertz) are respectively given by $(0.48,5.28)$,  $(0.577,3.88)$, $(0.63 ,3.17)$, $(0.701,2.30)$, and $(0.764 ,1.61)$. We fixed the flow rate of the current with a lower value than the one of the Vancouver experiments ($q=0.045 m^2 \, s^{-1}$) because it allows us to create currents with higher Froude numbers by diminishing the height without generating turbulence since the corresponding Reynolds number is lower in our experiments. It should be noticed that, although the values of $F_{\rm max}$ for our five flows differ from each other by only $50 \%$, $\om_{\rm min}$ varies from $5.28$ to $1.61$ Hz. To display the common properties of the transmission coefficient $\tilde{A}_\om$ in spite of this wide range, in \fig{fig:numpred}, we plot $\tilde{A}_\om$ as a function of $\om - \ommin$. 

We first notice that \eq{eq:stand4ode} and \eq{eq:neweq} both predict that $\tilde{A}_\om \ll 1$ for $\om \gg \ommin$, which means that the blocking is essentially complete. We also notice that $\tilde{A}_\om$ is close to unity for $\om \ll \ommin$, so that the incoming wave is essentially transmitted. Interestingly, for each flow, the transmission coefficient displays a sharp decrease in a narrow frequency region which is centered around the corresponding $\ommin$. With more precision, we found that the slope of $\tilde{A}_\om$ where $\tilde{A}_\om=0.5$ is close to $0.5 s$ for the standard method and $0.7 s$ for the refined one. As a result, the transition between the two regimes occurs within a narrow interval of order $1$ Hz around $\ommin$. Since $\ommin$ varies by a factor of three for the five flows we considered, these results are nontrivial. Indeed, the slope could have significantly varied, and the narrow intervals could have been centered on a different frequency than $\ommin$ (which is computed using the WKB approximation). The aim of the next subsections is to experimentally verify these two properties. 

It should be noticed that even though the predictions derived from \eq{eq:stand4ode} and \eq{eq:neweq} are rather similar, several differences should be noticed. First, in the upper panel the curves stop for a smaller value of $\om$. This is because the value of $\om_{\rm max}$ computed using the naive fourth-order expansion of the dispersion relation is significantly smaller than the exact one. Second, on the lower panel we observe that the curves obtained are neatly superimposed for $\om > \ommin$, and that their slopes are discontinuous at $\om = \ommin$. These seem to be artifacts of \eq{eq:neweq} when applied to the present obstacle.

\subsection{Experimental results}
\label{sub:expres}

\begin{figure}
\begin{center}
\includegraphics[scale=1.0]{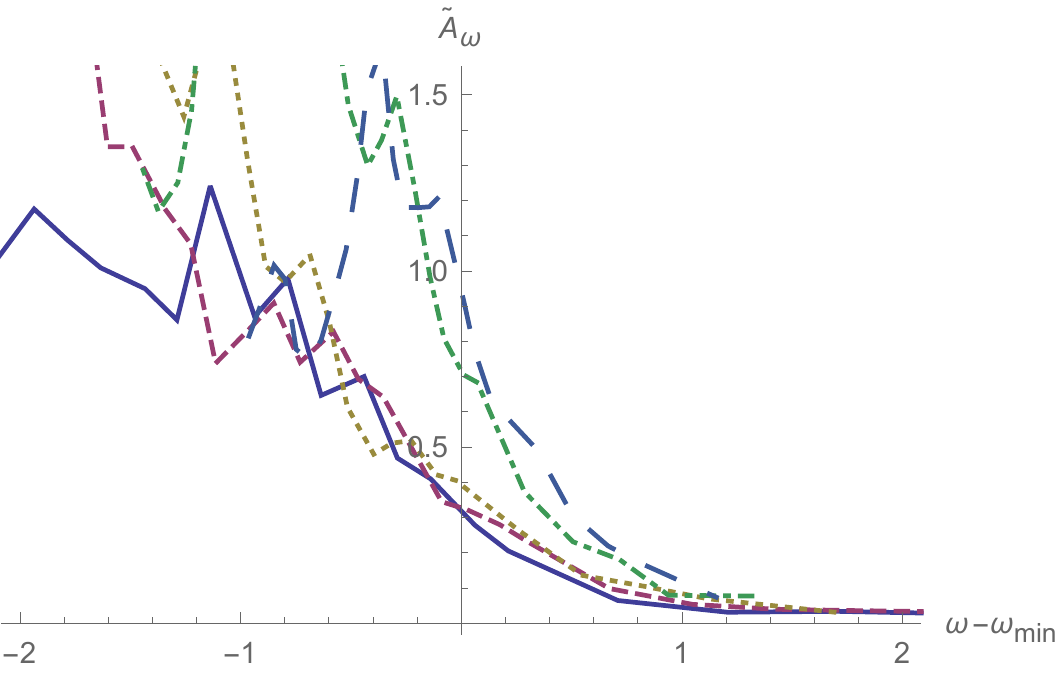}
\caption{Measured transmission coefficients as a function of $\om - \ommin$, for the five flows already considered in \fig{fig:numpred}. As explained in the text, the peaks at low frequencies seem to be due to multiple reflections on both ends of the flume.} \label{fig:superimposed}
\end{center}
\end{figure}

We measured the height of the free surface as a function of time using the setup described in
Sec.~\ref{subsection_experim_setup}, with two sensors in the downstream region, and two in the upstream one. The amplitude of the incident wave was measured in the downstream side of the obstacle and found to be close to $1.6$ mm. One sensor would have been sufficient, but the second one allowed us to check that the two measurements of the peak to peak amplitudes were in agreement. When the transmission was significant, these measurements were not accurate as the waves reflected at the end of the flume interfered with the incident one. We thus extrapolated the amplitude of $1.6$ mm found for all larger values of $\om$ to these frequencies. This is justified since the amplitude was found to be independent on $\om$ for all frequencies $\om > \ommin$.

In the upstream region the situation is more complicated because the transmitted wave is (partially) reflected by the honeycomb in the convergent chamber. Hence, the sensors on the upstream side of the obstacle in fact measured the {\it superposition} of the transmitted wave and the wave reflected at the entrance of the channel. One thus anticipates an interference pattern, as the amplitude measured by a sensor can oscillate between zero and (about) twice the amplitude one would have obtained without the reflection. In fact, the sharpness of the observed peaks indicate that there should be multiple scattering (on both ends of the flume). These should engender an interference pattern that we did not try to determine. It should be also possible to measure it by varying the distance between the upstream sensors. However, this would be very difficult since this distance, $0.4$ m,  is small with respect to the long wavelength of the transmitted wave. In practice, to reduce the noise and the possibility of measuring a null amplitude if one sensor would be at a node of the interference pattern
(at least for not too long a wavelength), we took the mean value of the signals measured by the two upstream sensors separated by a distance taken as large as possible given the mechanical constraints. To extract the signal associated with the transmitted wave from this mean value, we performed a Fourier transform and observed a narrow signal at the frequency of the wave generator, so there is no ambiguity in measuring the amplitude of the wave in the upstream region. 

Notice finally that since the water heights are the same on both sides of the obstacle, the ratio of the transmitted and incident amplitudes should give directly (ignoring the above discussed reflections) the transmission coefficient $\tilde{A}_\om$ of \eq{eq:Bsub} which relates unit norm waves. In \fig{fig:superimposed}, we present the experimental results for the same set of flows as that used in \fig{fig:numpred}. We clearly observe the three main theoretical predictions. First, for $\om$ significantly larger than $\ommin$, $\tilde{A}_\om$ is very small. Second, $\tilde{A}_\om$ becomes of order 1 for $\om < \ommin$. Third, more importantly, for the five flows, the transition occurs in a narrow interval of $\sim 2$ Hz which, moreover, is centered close to the value of $\ommin$ of the corresponding flow.

We also observe that the curves are relatively smooth for $\om > \ommin$, but show strong peaks for $\om < \ommin$. As these peaks are sharp and make $\tilde{A}_\om$ go above 1, we believe they are due to the finite size of the flume. As discussed above, reflections on both ends of the flume are expected to significantly affect the measurements of the amplitude of the transmitted wave.

\subsection{Comparison of numerical and experimental results}

\begin{figure}
\begin{center}
\includegraphics[scale=1.0]{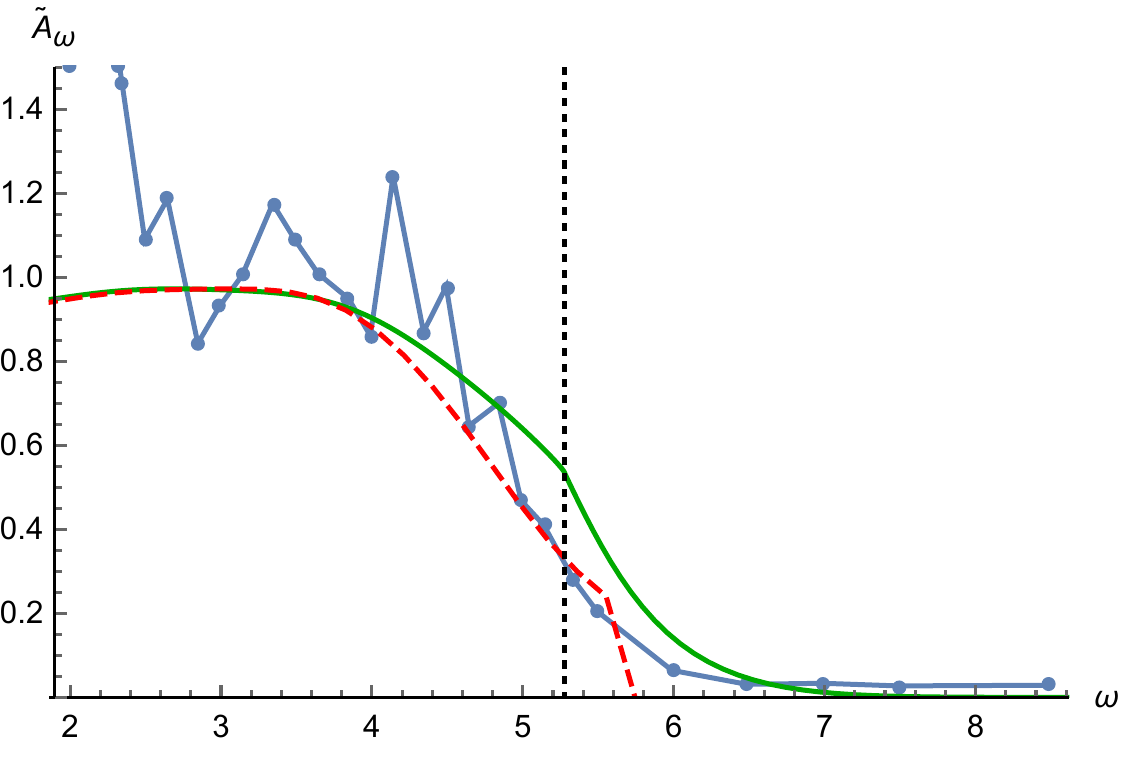}
\end{center}
\caption{Plot of the transmission coefficient $\tilde{A}_\om$ vs the angular frequency $\om$ (in hertz) for a flow over the obstacle. The asymptotic water depth is $h_d = 0.181 \; \text{m}$ and the flow is $q = 0.0276 m^2.s^{-1}$. The green, solid curve shows the numerical prediction obtained using the method based on \eq{eq:neweq}, and the red, dashed one shows results from a standard fourth-order expansion of the dispersion relation around $k=0$. Dots are the experimental data. The dotted vertical line indicates $\om_{\rm min}$.}\label{fig:firstcomp} 
\end{figure}
To illustrate the quality of the correspondence between theoretical predictions and experimental data, in \fig{fig:firstcomp}, we plot $\tilde{A}_\om$ as a function of $\om - \ommin$ for the flow with the largest asymptotic water depth. Overall, we observe a good agreement between the three curves. In particular, they show the same strong decrease of $\tilde{A}_\om$ near $\ommin$. With more details, for small values of $\om$, we see that the two numerical methods agree very well with each other. They both predict that the transmission coefficient goes to 1 in the small-frequency limit $\om \to 0$. In this domain the agreement with experimental data is rather poor. We believe this is due the reflection of the transmitted wave mentioned in subsection~\ref{sub:expres}. When approaching $\om = \ommin$, the predictions of the two numerical methods show a non-negligible difference. Interestingly, experimental data show a better agreement with the ``naive'' one based on \eq{eq:stand4ode}. However, given the expected uncertainties,  this could be fortuitous. 

\begin{figure}
\begin{center}
\includegraphics[scale=1.0]{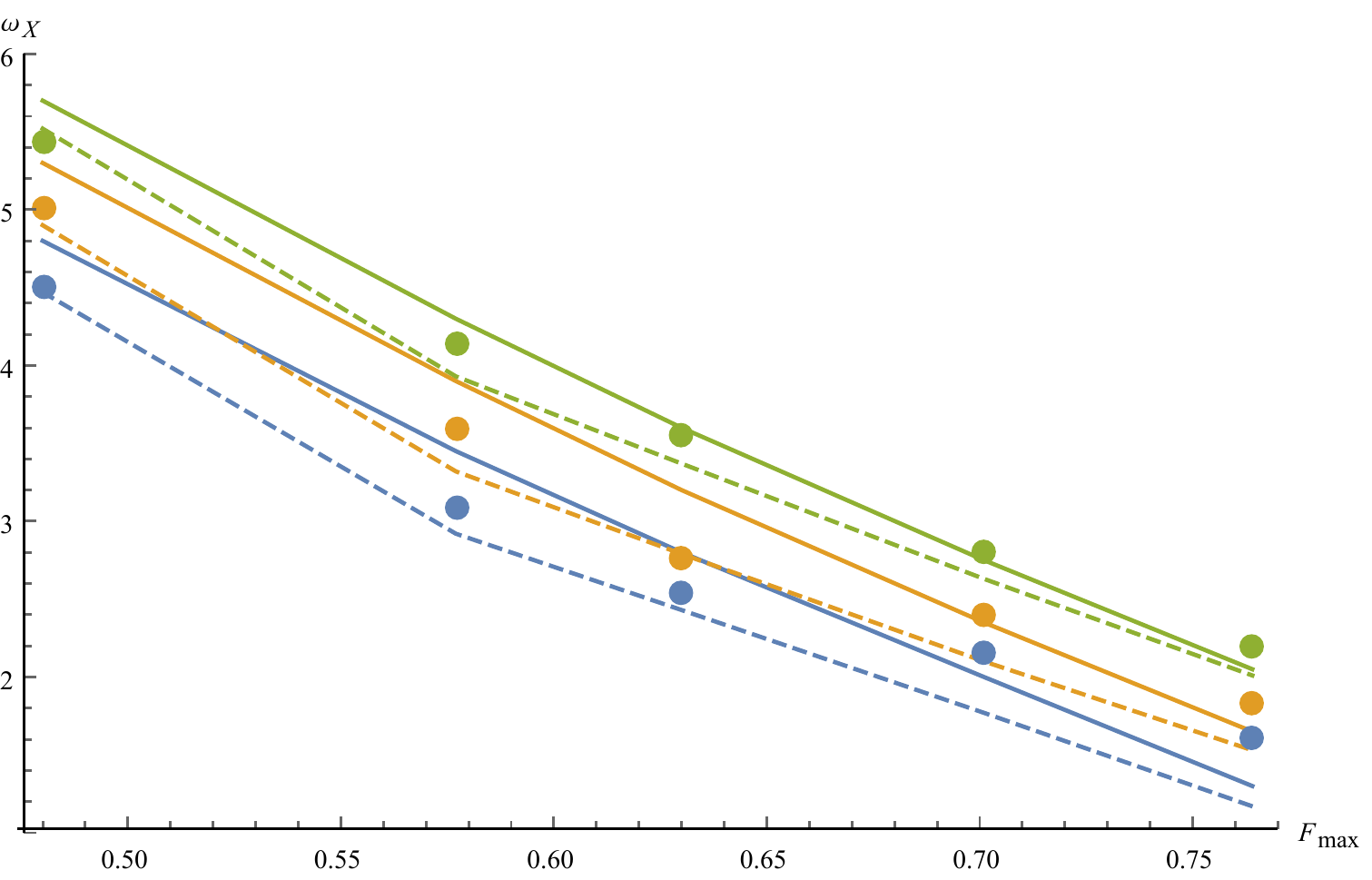}
\end{center}
\caption{Plot of the frequencies $\om_X$ where $\tilde{A}_\om$ reaches the value $X$, for $X = 1/\sqrt{2}$ (green), $1/2$ (orange), and $1/4$ (blue). Plain lines are obtained using the refined wave equation of \eq{eq:neweq}, dashed ones are obtained using the standard quartic wave equation of \eq{eq:stand4ode}, 
and dots are the experimental data.}\label{fig:mainResults}
\end{figure}
 To compare the numerical and experimental results more quantitatively, we show in \fig{fig:mainResults} the theoretical and  experimental values of the angular frequencies $\om_X$ for which $\tilde{A}_\om$ reaches $X=1/\sqrt{2}$, $1/2$, and $1/4$, for the five flows previously described. We notice that the agreement between numerical and experimental data is quite good. Indeed, the errors are of the order of $10 \%$ or smaller, except for the two last flows for $X = 1/\sqrt{2}$ where the contribution of spurious narrow peaks is large. We find the agreement is best for $X=1/4$, as expected since the experimental data are less noisy for higher frequencies. The present agreement between theory and experimental data with relative errors of about $10 \%$ is the main result of this work.

\begin{figure}
\begin{center}
\includegraphics[scale=1.0]{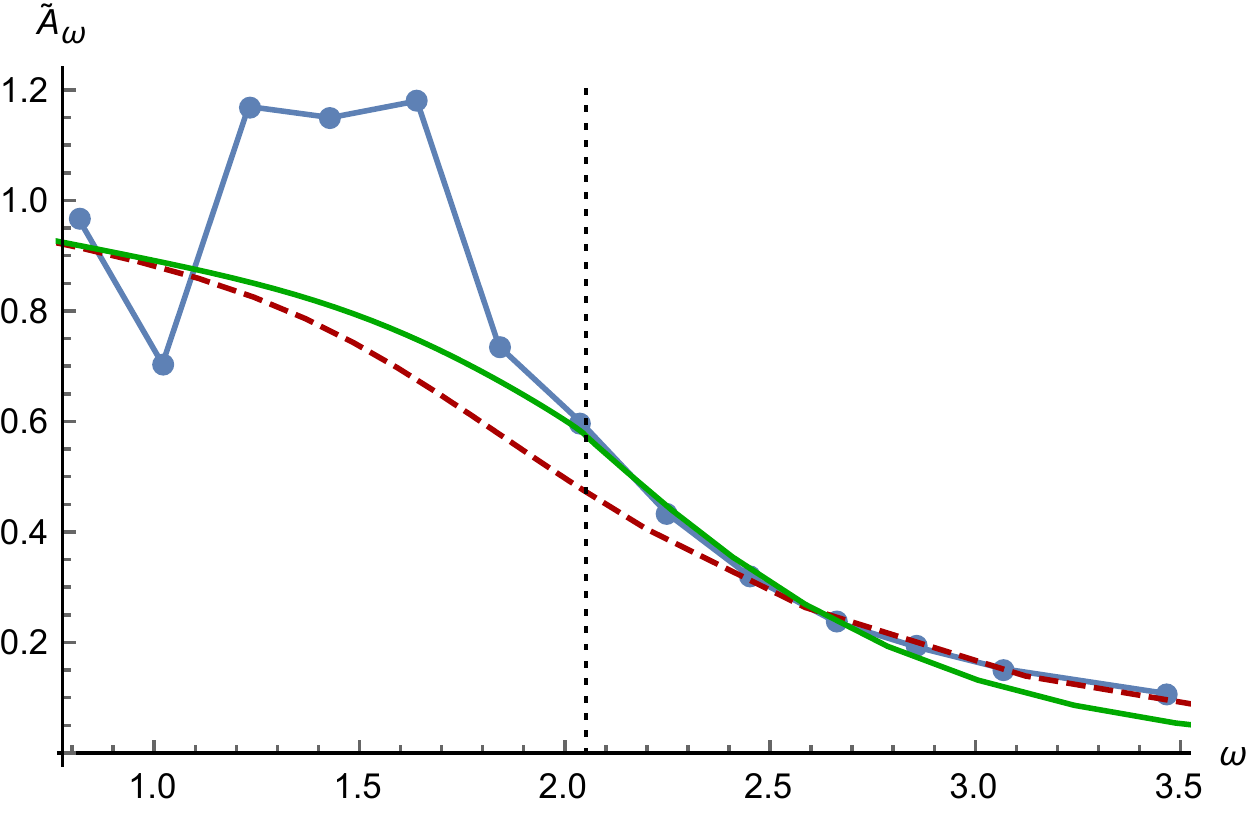}
\end{center}
\caption{Plot of the transmission coefficient $\tilde{A}_\om$ as a function of $\om$ for an asymptotic water depth of $0.194$ m and a current flow rate $q=0.045 m^2 \cdot s^{-1}$. The green plain curve is obtained using the refined numerical method, the dashed red one is obtained using the standard fourth-order expansion of the dispersion relation, and dots are the experimental data. The dotted vertical line indicates the value of the angular frequency $\ommin = 2.05$ Hz obtained from the simplified analysis of the background flow of Appendix A.}
\label{compVancouver}
\end{figure}

\subsection{Application to the Vancouver experimental setup} 

To conclude this study, we consider the settings of the Vancouver experiment~\cite{Unruh2010}, namely with an asymptotic water depth of $0.194$ m and a current flow rate of $q=0.045 m^2 \cdot s^{-1}$. We recall that these have been chosen to detect the analogue Hawking radiation by observing the ratio $R_\omega = \left\lvert \beta_\om \right\rvert^2 / \left\lvert \alpha_\om \right\rvert^2$ of the coefficients of the dispersive waves of opposite norms in \eq{eq:Bsub}. Before discussing the behavior of the transmission coefficient $\tilde A_\om$, we point out that we also measured $R_\omega$ using techniques similar to those used in Vancouver. In agreement with what was reported~\cite{Unruh2010}, we found that $\ln R_\omega $ is approximatively linear in $\omega$ with a similar value of the slope. This indicates that the forthcoming analysis, which concerns a coefficient which was not studied in Ref.~\cite{Unruh2010}, is compatible with the results of this work.

Given the asymptotic water depth and the flow rate, we obtain that the maximum value of $F$ is $F_{\rm max} \approx 0.7$, and that the critical frequency of \eq{eq:ommin} is $\om_{\rm min} \approx 2$Hz. ~\footnote{These values of $\ommin$ and $F_{\rm max}$ are computed using the measured minimal value of $h$, and under the approximations mentioned at the beginning of Sec.~\ref{sub:gen}. This minimum value of $h$ closely matches that obtained from the analysis presented in Appendix. A, as well as the refined treatment presented in Ref.~\cite{Germain4} and applied in the Appendix A of~\cite{MP14}. However it must be noted that measurements of the wavelength of the undulation in the downstream region seems to indicate that the velocity at the surface is larger than expected by about 10\%. We believe that the origin of the discrepancy is mainly due to the neglect of the vorticity of the flow near the bottom. This could notably affect the values of $\ommin$ and $F_{\rm max}$. We hope to shortly clarify this point using more precise data obtained from particle image velocimetry.} In \fig{compVancouver} we present our numerical and experimental results. As in the previous cases, we observe a relatively good agreement between the theoretical and the observational curves. In particular, we see that the transmission coefficient $\tilde{A}_\om$ is larger than $1/2$ for $\om < \ommin$. More precisely, we see that $|\tilde{A}_\om|^2 = 1/16$ is reached for $\om \approx 2.6$ Hz. Accepting about $6 \%$ of relative errors, this means that the contribution of the transmission coefficient to the unitarity relation \eq{eq:unit} can be neglected only for relatively high frequencies. In fact, for five of the nine experimental points of the Fig.~5 of Ref.~\cite{Unruh2010}, $|\tilde{A}_\om|^2$ is larger than $1/16$. Because of this, the unitarity relation \eq{eq:unit} differs from the standard one which reads $\left\lvert \alpha_\om \right\rvert^2 - \left\lvert \beta_\om \right\rvert^2 = 1$.
Hence, \eq{eq:unit} cannot be used to support the interpretation of the linear behavior of $ln R_\om$ as an indication of the Planckianity of the spectrum. We refer to the detailed analysis of Ref.~\cite{MP14} for a discussion of these matters, and the conditions needed to have a clear relation with the Hawking effect. In particular, it is shown that the subcritical character of the flow implies that $|\beta_\om|^2$ of \eq{eq:Bsub} remains much smaller than 1, typically of order $10^{-5}$, instead of diverging as $1/ \om$, as it is found in transcritical flows where the correspondence with black hole physics is much clearer. 

We also note that the good agreement between the theoretical curves and the
observational data concerning the transmission coefficient $\tilde A_\om$
is a strong indication that the predictions of Ref.~\cite{MP14} concerning the
other three coefficients of \eq{eq:Bsub}; see Fig.~9 left panel, are also in good agreement with the physics. We believe this is the second important outcome of the present paper.

\section{Conclusions} 
\label{Con}

We studied the transmission coefficient of counterpropagating shallow-water waves over a localized obstacle, both theoretically and in an experimental setup. For a given obstacle and a given current, we considered a series of five subcritical flows obtained by varying the asymptotic value of the water height. The maximal value of the Froude number $F_{\rm max}$ reached on top of the obstacle varied from $0.48$ to $0.764$, and the corresponding critical frequency $\ommin$ varied from $5.28$ to $1.61$Hz. For each flow, the $\sim 25$ values of the probed frequencies ranged from $0.4 \, \ommin$ to $1.7 \, \ommin$. We found a clear quantitative agreement between the theoretical predictions and the experimental data. 

The main property of the transmission coefficient which is common to all flows is the following: for an increasing value of the frequency, the coefficient sharply decreases from essentially 1 (transmission) to zero (blocking) in a narrow frequency interval (of $\pm 1$ Hz) centered around the critical frequency $\ommin$. In addition, we found that the slope in this interval hardly depends on the value of $\ommin$. (This is correct at least for low Froude numbers. For the two highest values, the contribution of narrow peaks associated with multiple reflections seem to contaminate the slope.) To complete our analysis, we also considered the transmission coefficient in the flow used in the Vancouver experiment. We basically recovered the same properties as the above ones. These imply that the transmission coefficient cannot be neglected for about the lower half frequency range that was probed. 

At this point, it should be underlined that the theoretical analysis~\cite{MP14} indicates that a clear correspondence with the Hawking effect, namely a mode mixing coefficient $|\beta_\om|^2$ of \eqref{eq:Bsub} growing as $1/\om$ in a sufficiently large frequency range, can be found only if the transmission coefficient is sufficiently small. It is therefore important to reduce the transmission at low frequency, i.e. to improve wave blocking. To this end it is necessary to increase the maximal value of the Froude number. Indeed, as shown in Eqs.~(\ref{eq:omm},\ref{eq:ommin}), $\ommin$ decreases as $(1 - F_{\rm max}^2)^{3/2}$.  In future experiments, one should thus try to work with significantly higher values of $F_{\rm max}$ than those used in Refs.~\cite{Nice,Unruh2010}.

In the future, we also hope to be able to measure the scattering coefficients of the four outgoing waves, and to validate the predictions obtained by using the improved quartic equation we introduced in Sec.~\ref{sub:num}. On the numerical side, it would be interesting to see how the results are affected when using a scheme which approximates the exact dispersion relation more closely. It could even be possible to take the full dispersion relation into account using the technique described in Ref.~\cite{Scott_numerical_method_1}. The description of the background flow should also be improved if one aims at getting relative errors $\ll 10\%$. This could be achieved by using a low gradient approximation if the Froude number does not change significantly along the flow. It would also be important to study the consequences of the multiple scattering on both ends of the flume, and those of the undulation, as was recently done in Ref.~\cite{Busch:2014hla} in the context of atomic Bose-Einstein condensates. 

\acknowledgements

We would like to thank Romain Bellanger, Jean-Marc Mougenot and Patrick Braud for their help in the course of the experiments. Yvan Jolit and Jean-Marc Mougenot designed the wave-maker. We thank Scott Robertson and Iacopo Carusotto for interesting comments. We also thank Bill Unruh and Silke Weinfurtner for many interesting discussions. This work has benefited from the following funding: CNRS Interdisciplinary Mission in 2013, ACI "Wave-current Interaction" from the University of Poitiers in 2013, and CNRS Interdisciplinary PEPS PTI "DEMRATNOS" in 2014.

\normalsize

\appendix

\section{Theoretical description of background flows} 
\label{app:bf} 

We briefly explain the procedure we adopt to get an approximate description of the stationary background flows over the obstacle we used in the experiment. Because the Froude number remains sufficiently smaller than 1, it is legitimate to use Bernoulli equation, and to neglect the vertical component of the flow velocity. These approximations will be justified {\it a posteriori} using a self-consistency criterion, and ultimately by the relatively good agreement with experimental data. 

Only stationary longitudinal two-dimensional flows are here considered, following ref.~\cite{Germain4}. The longitudinal coordinate is called $x$, and the vertical one is called $y$. We denote as $y =p(x)$ the height of the obstacle, assumed to vanish asymptotically on both sides, $P$ the local pressure, and $\rho$ the (constant) density of the fluid. The Bernoulli equation gives~\cite{LLhydro}
\be 
\vec{\nabla} \lp \frac{1}{2} {\mathbf{v}}^2 + g y + \frac{P}{\rho} \rp = 0.
\ee  
Neglecting the surface tension (so that $P$ is a constant along the free surface) and the vertical component of the velocity, one obtains
\be \label{eq:nB}
\frac{1}{2} \frac{q^2}{h(x)^2} + g (h(x)+p(x)) = \frac{1}{2} \frac{q^2}{h_0^2} + g (h_0+p_0),
\ee
where $h_0$ and $p_0$ are the water depth and obstacle height at a given point $x=x_0$. In practice we chose $x_0$ to be a point where $p$ is maximal, so that the shape of the free surface is close to the actual one where the Froude number reaches its maximum. It is easily shown that \eq{eq:nB} has two positive solutions in $h$ provided
\be \label{eq:condnB}
\frac{3}{2} \lp g q \rp^{2/3} + g p < \frac{1}{2} \lp \frac{q}{h_0} \rp^2+g (h_0 + p_0).
\ee
The smallest one is supercritical $v > c$, while the largest one is subcritical $v < c$. If \eq{eq:condnB} is not satisfied, \eq{eq:nB} has no positive solution. For the parameters we shall consider, \eq{eq:nB} is always satisfied. Moreover, we only consider subcritical flows, so we will always choose the largest value of $h(x)$, solution of \eq{eq:nB}. 

One can now check the hypothesis that $v_y \ll v_x$. To this end, since the free surface is a streamline, we have
\be 
\frac{v_y}{v_x} = \frac{d}{d x} \lp h+b \rp.
\ee
An easy calculation using \eq{eq:nB} shows that this is equal to
\be 
\frac{v_y}{v_x} = \frac{F^2}{F^2-1} \frac{d b}{d x},
\ee
which can be easily evaluated given the shape of the obstacle. For the flows we shall consider this quantity is always smaller than $10 \%$ in absolute value. So, neglecting $v_y$ before $v_x$ is justified. One possible loophole in this argument is the appearance of an undulation, i.e., a zero-frequency mode with nonvanishing wave vector, which is not described by~\eq{eq:nB} but which was observed in the laboratory. Even if its amplitude is small,  it could in principle have an important effect as it extends in the whole downstream region. However, the preliminary analysis of~\cite{MP14} indicates that its effect on the transmission coefficient is rather small.

\bibliographystyle{apsrev4-1}
\bibliography{biblio}

\end{document}